\documentclass[aps,prl,twocolumn,superscriptaddress,showpacs]{revtex4}
\usepackage{graphicx}

\begin{document}   



\title{Electronic Aharonov-Bohm Effect Induced by Quantum Vibrations}
\author{R. I. Shekhter}
\affiliation{Department of Physics, G\"{o}teborg University,
SE-412 96 G\"{o}teborg, Sweden}
\author{L. Y. Gorelik}
\email{gorelik@fy.chalmers.se} \affiliation{Department of Applied
Physics, Chalmers University of Technology, SE-412 96
G\"{o}teborg, Sweden}
\author{L. I. Glazman}
\affiliation{W.I. Fine Theoretical Physics Institute,
University of Minnesota, Minneapolis, MN 55455}
\author{M. Jonson}
\affiliation{Department of Physics, G\"{o}teborg University,
SE-412 96 G\"{o}teborg, Sweden}
\date{\today}


\begin{abstract}
Mechanical displacements of a nanoelectromechanical system (NEMS)
shift the electron trajectories and hence perturb phase coherent
charge transport through the device. We show theoretically that in
the presence of a magnetic feld such quantum-coherent
displacements may give rise to an Aharonov-Bohm-type of effect. In
particular, we demonstrate that quantum vibrations of a suspended
carbon nanotube result in a positive nanotube magnetoresistance,
which decreases slowly with the increase of temperature. This
effect may enable one to detect quantum displacement fluctuations
of a nanomechanical device.

\end{abstract}

 \maketitle

Following the ubiquitous trend of downsizing devices,
micromechanical systems (MEMS) are today evolving into
nanoelectromechanical systems (NEMS) rapidly approaching the
limits set by the laws of quantum mechanics \cite{PhysicsToday}.
The ultimate potential for nanomechanical devices is governed by
the ability to detect the NEMS motional response to various
external stimuli. In the quantum regime of operation optical and
other sensing methods used in MEMS are not practical and one has
turned instead to various mesocopic sensing devices. Much work,
including fundamental research, still needs to be done. For
example, even though one has recently been able to detect flexural
vibrations of a SiN beam resonator with the amazing sensitivity of
10$^{-13}$~m using a radio-frequency single-electron transistor
\cite{LaHaye}, this is still about 6 times the quantum limit set
by the amplitude of the zero-point oscillations of the beam.

In this Letter we propose a different approach to sensing
ultrasmall quantum vibrations of a beam --- we have a suspended
carbon nanotube in mind --- and show that coherent tube vibrations
can induce an effectively multi-connected electron path through
the tube. Through an Aharonov-Bohm-type effect this in turn gives
rise to a negative magneto-conductance that can be detected.
We propose that this is but one example of how employing quantum
coherence in both the electronic and mechanical degrees of freedom
may lead to new functionality and novel applications.

Fundamental research on NEMS in the quantum coherent regime can
profit from analogies with mesoscopic phenomena in confined
structures, where the conductance may depend on quantum
interference between electron waves. It is, {\em e.g.}, well known
that the conductance of a mesoscopic sample changes if a single
impurity is displaced a small distance and that varying a magnetic
field leaves a sample-specific ``magnetic fingerprint" conductance
pattern. We will show that a similar effect can be achieved by the
mechanical displacement corresponding to quantum vibrations of a
suspended carbon nanotube in the presence of a magnetic field. To
this end, consider first the structure shown in Fig.~1(b). Here a
beam of electrons can pass through two openings in an otherwise
opaque screen. Interference between the quantum amplitudes for
going through one or the other of the two holes will determine the
probability for electrons to hit the detector. Obviously no such
interference effect is expected with only one hole, as in
Fig.~1(a). Even if the position of the hole in the screen were to
move, no interference would occur if the trajectory of the hole is
classically well defined. In every instant, only one classical
trajectory would be relevant.
\begin{figure}
  \label{Fig0}
  \includegraphics[scale=0.39]{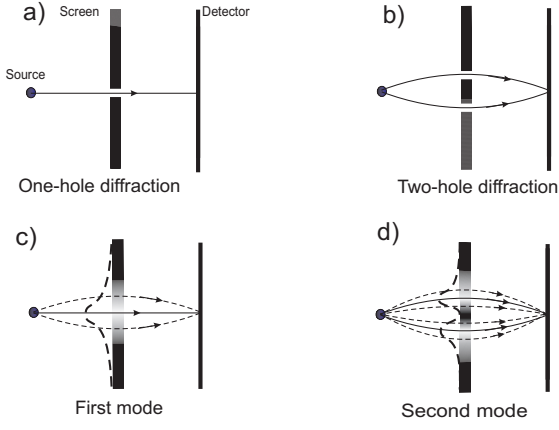}
  \vspace{-1 mm}
  \caption{ \label{fig:case22} Illustration of
  multi-connectivity in phase coherent electron transport by analogy
  with electron diffraction. In panel (b) electrons pass through two
  holes in a screen before hitting the detector.
  Unlike in panel (a), where the screen has only one hole, interference
  between the amplitudes
  of the two quasi-classical electron paths (solid arrows)
  determines the probability for electrons to hit the detector.
  In panels (c) and (d) electrons pass through a ``quantum" hole,
  whose position is determined by a wave function with zero (c) and one (d) node.
  The fuzziness of the hole effectively creates different ``quantum"
  electron paths of high (filled arrows) and low (dashed arrows)
  probability amplitude. In the text this analogy is shown to be valid
  for electrons tunneling through a suspended carbon nanotube in situations where
  only a few vibration states are allowed to couple to the
  electrons.}
  \vspace{-5 mm}
\end{figure}
This situation changes qualitatively if the hole motion is in the
quantum regime, where its position cannot be precisely defined and
``quantum alternative" paths for the electrons through the screen
appear. Figures~1(c) and (d) show two examples of such a ``quantum
geometry", giving rise to electron paths through the hole that are
multiply connected. Such an intimate, coherent
nanoelectromechanical coupling between electronic and mechanical
degrees of freedom leads to characteristic quantum interference
among the electron paths and in particular to an
Aharonov-Bohm-type of effect in the presence of an external
magnetic field.

Due to their low mass and unique mechanical and electronic
properties, single-wall carbon nanotubes (SWNTs) offer perhaps the
best possibility for studying quantum nanoelectromechanical
phenomena \cite{LeRoy,OurNanoLett}. Figure~2 shows a sketch of the
system we have in mind to achieve coherent coupling between
quantum electronic transport and quantum flexural vibrations of a
nanotube: a free-hanging SWNT, doubly clamped to two metallic
leads and subject to a transverse magnetic field, $H$. This system
is described by the Hamiltonian,
\begin{equation}\label{Hin}
 \hat{ H}= \hat{ H}_{\rm leads}+\hat{ H}_{\rm el}+\hat{ H}_{\rm mech}+\hat{ H}_{\rm
 tunn}\,,
\end{equation}
where the first term,
\begin{equation}\label{Ha}
\hat{ H}_{\rm leads} =
\sum_{k}\varepsilon_{\ell,k}\hat{a}^{\dag}_{\ell,k}\hat{a}_{\ell,k}
+ \sum_{k}\varepsilon_{{\rm r},k}\hat{a}^{\dag}_{r,k}
\hat{a}_{{\rm r},k}\,,
\end{equation}
models electrons in states $k$ in the left ($\ell$) and right
($r$) leads and $\hat{a}^{\dag}_{\ell/{\rm r},k}$
[$\hat{a}_{\ell/{\rm r},k}$] is the corresponding creation
[annihilation] operator. The second term,
\begin{eqnarray}\label{Ht}
\hat{ H}_{\rm el}&=&\int d^{3}\vec{r} \left\{
-\frac{\hbar^{2}}{2m}\hat{\psi}^{\dag}(\vec{r})\left(\frac{\partial}{\partial
\vec{r}}- \frac{ie}{c\hbar}\vec{A}(\vec{r})
\right)^{2}\hat{\psi}(\vec{r})+ \right. \nonumber \\
 &&+\left. U(y-\hat{u}(x),z)\hat{\psi}^{\dag}(\vec{r})\hat{\psi}(\vec{r})\right\}\,,
\end{eqnarray}
describes the SWNT electrons, 
confined in the transverse direction by a potential $U(y,z)$ that
depends on the deflection $u(x)$ of the tube (in the
$y$-direction). The operator $\hat{\psi}^{\dag}(\vec{r})$
[$\hat{\psi}(\vec{r})$] creates [annihilates] an electron at
$\vec{r}=(x,y,z)$; $\{\hat{\psi}^{\dag}(\vec{r}),
\hat{\psi}(\vec{r}')\}=\delta(\vec{r}-\vec{r}')$ and
$\vec{A}(\vec{r})=(-Hy,0,0)$.

The bending of the tube is modelled by the third term in the
Hamiltonian (\ref{Hin}) as
\begin{equation}
 \hat{ H}_{\rm mech}=\int^{L/2}_{-L/2} dx \left\{\frac{1}{2\rho}
 \hat{\pi}^{2}(x)+\frac{EI}{2}\left(\frac{\partial^{2} \hat{u}(x)}{\partial
 x^{2}}\right)^{2}\right\}\,.
\end{equation}
Here $\hat{\pi}(x)$ is the momentum density operator conjugate
with the deflection field operator $\hat{u}(x)$, {\em i.e.}
$[\hat{u}(x),\hat{\pi}(x')]=i\hbar\delta(x-x')$), $\rho$ is the
linear mass density of the SWNT, $I$ is its area moment of inertia
and $E$ is the Young's modulus. The tube is doubly clamped, which
gives the boundary conditions $u(x)=0$ and $u'(x)=0$ for $|x|\geq
L/2$.
\begin{figure}
  \label{Fig1}
  \includegraphics[scale=0.5]{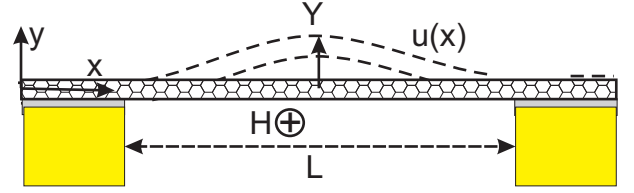}
  \vspace{-1 mm}
  \caption{ \label{fig:case22} Nanoelectromechanical system
  proposed to show the coherent coupling between quantum electron transport
  and quantum flexural vibrations discussed in the text. Electrons
  tunneling through a doubly clamped single-wall carbon
  nanotube (SWNT) excite quantized vibrations of the SWNT in the
  presence of a magnetic field, $H$. The resulting effective multi-connectivity
  of the system leads to a negative magneto-conductance
  (see text).}
  \vspace{-5 mm}
\end{figure}

The tunneling Hamiltonian, $\hat H_{\rm
tunn}=\hat{T}_{\ell}+\hat{T}_{\rm r}$, where
\begin{equation}\label{Ht}
\hat{T}_{\ell/\rm r}=\sum_{k} \int d\vec{r}\, {\cal T}_{\ell/\rm
r}(\vec{r},k) \hat{\psi}^{\dag}(\vec{r})\hat{a}_{\ell/{\rm r},k}+
h.c.\,,
\end{equation}
and ${\cal T}_{\ell/\rm r}(\vec{r},k)$ are overlap integrals,
describes how electron tunneling couples the SWNT and the two
leads.

In order to proceed it is convenient to make the unitary
transformation $e^{i\hat{S}}\hat{ H}e^{-i\hat{S}}$, with
\begin{eqnarray}
\hat{S}&=&-i\int
d^{3}\vec{r}\left\{\hat{u}(x)\hat{\psi}^{\dag}(\vec{r})\frac{\partial
\hat{\psi}(\vec{r})}{\partial y}+\right. \nonumber \\
&&\left.i\frac{eH}{\hbar c}
\left(\int_{0}^{x}dx'\hat{u}(x')\right)\hat{\psi}^{\dag}(\vec{r})\hat{\psi}(\vec{r})\right\}\,.
\nonumber
\end{eqnarray}
Here the first term produces a coordinate transformation to the
nanotube reference frame, while the second generates a gauge
transformation that eliminates the vector potential from the
Hamiltonian $\hat H_{\rm el}$. Furthermore, since the transverse
electron motion in the SWNT is strongly quantized it may be
decoupled from the longitudinal motion by letting
$\hat{\psi}(\vec{r})=\Psi(y,z)\hat{\psi}(x)$. Here
$\Psi(y,z)=\Psi(\vec{r_{\rm t}})$ is the wave function
corresponding to a transverse quantized energy level $E_{\rm t}$.
As a result the terms $\hat{ H}_{\rm el}$, $\hat{ H}_{\rm mech}$
and $\hat{T}_{l/r}$ in the Hamiltonian (1) simplify to
\begin{eqnarray}\label{Ht2}
\hat{ H}_{\rm el}&=&\int dx \,
\hat{\psi}^{\dag}(x)\left(-\frac{\hbar^{2}}{2m}\frac{\partial^{2}}{\partial
x^{2}}+\varepsilon_{\rm t} \right) \hat{\psi}(x)\nonumber \\
\hat{ H}_{\rm mech}&=&\int^{L/2}_{-L/2} dx
\left\{\frac{1}{2\rho}\left(\hat{\pi}(x)-\frac{eH}{c}\int_{0}^{x}dx'\hat{\psi}^{\dag}(x)\hat{\psi}(x)\right)^{2}\right.\nonumber\\
&&+\left.\frac{EI}{2}\left(\frac{\partial^{2}
 \hat{u}(x)}{\partial
 x^{2}}\right)^{2}\right\}\\
\hat{T}_{l/r}&=&T {\rm exp}\left\{i\frac{e H}{\hbar c}\int^{ \mp
L/2}_{0}dx\hat{u}(x)\right\}\times\nonumber \\
&&\sum_{k} \int dx T_{\rm \ell/ r} (x,k)
\hat{\psi}^{\dag}(x)\hat{a}_{k,\rm \ell/ r}+ h.c \nonumber \,,
\end{eqnarray}
where $T_{\rm \ell/ r} (x,k) = \int d\vec{r_{\rm t}}{\cal
T}_{\ell/\rm r}(\vec{r},k)|\Psi(\vec{r_{\rm t}})|^{2}$ and
\cite{et} $\varepsilon_{\rm t} =E_{\rm
t}+(e^{2}H^{2}/2mc^{2})\int\int d\vec{r_{\rm t}}\,
y^{2}|\Psi(\vec{r_{\rm t}})|^{2}$.

By analogy one may think of the elementary excitations created by
$\hat{\psi}^{\dag}(x)$ in the transformed Hamiltonian (\ref{Ht2})
as polarns. It is important that due to the quantum vibrations of
the nanotube the wavefunction of this polaronic state is extended
in the direction perpendicular to the the tube axis. These
``swinging states" serve as intermediate states for electrons
tunneling through the vibrating SWNT. The quantum phase acquired
by the tunneling electrons depends on magnetic field and below we
will show that a nanomechanically induced magneto-conductance may
follow.

So far our approach has been quite general, but here we will make
some approximations that allow us to find an analytical solution.
First, we consider the weak coupling limit, where the spacing
$\Delta E \simeq {\rm v}_{F}\hbar L_{0}$ of quantized energies
associated with the {\em longitudinal} motion of electrons along
the SWNT of total length $L_{0}$ is much larger than the level
broadening caused by the coupling to the continuous spectrum of
energy levels in the leads (${\rm v}_{F}$ is the Fermi velocity).
In this case only a few longitudinal states with energy
$\varepsilon_{n}$ are relevant. Further simplifications arise if
one restricts the mechanical SWNT dynamics to the fundamental
bending mode, which gives the most important contribution, and
expresses the corresponding deflection operator as
$\hat{u}(x)=Y_{0}u_{0}(x/L) (\hat{b}^{\dag}+\hat{b})/\sqrt{2}$,
where $u_{0}(\xi)$ is the normalized profile of the fundamental
bending mode, $Y_{0}=({\hbar^{2}L^{2}/\beta_{0}\rho EI})^{1/4}$ is
the amplitude of zero-point fluctuations in the fundamental mode,
$\beta_{0}$ is the dimensionless eigenvalue associated with this
mode and $\hat{b}^{\dag}$ [$\hat{b}$] is a boson operator that
creates [annihilates] one vibration quantum.

Next we consider a non-resonant case where no electron energy
level in the SWNT is close enough to the chemical potential $\mu$
in the leads to be involved in even inelastic electron tunneling
via a real state on the tube. This requires the criteria
$|\varepsilon_{n}+\varepsilon_{\rm t}-\mu|\gg \nu|T_{\ell/\rm
r}|^{2},\hbar \omega, eV$ to be fulfilled, where $\nu$ is the
electron density of states in the leads, $\omega
=(\beta_{0}IE/\rho)^{1/2}/L^{2}$ is the eigenfrequency of the
fundamental mode and $V$ is the applied bias voltage. Under such
conditions the coupling between electronic states in the left and
right leads occurs via virtual states on the tube and may be
described by an overlap integral. As a result one comes to a
tunneling Hamiltonian,
\begin{eqnarray}\label{H4}
\hat{ H}_{\rm eff}&=& \sum_{k;\sigma=\ell,{\rm
r}}\varepsilon_{\sigma,k}\hat{a}^{\dag}_{\sigma,k}\hat{a}_{\sigma,k}^{
}+ \hbar\omega \hat{b}^{\dag}\hat{b}+ \\
&&
 e^{i\phi(\hat{b}^{\dag}+\hat{b})}\sum_{k,k'}
T_{\rm eff}(k,k')\hat{a}_{{\rm r},k}^{\dag}\hat{a}_{{\ell},k'}^{
}+h.c. \nonumber \,,
\end{eqnarray}
that describes non-resonant charge transfer through the suspended
SWNT. In this equation $\phi = g4\pi Y_{0}LH/\Phi_{0}$ is a
dimensionless magnetic flux, $\Phi_{0}=hc/e$ is the flux quantum,
$g=(\pi/\sqrt{2})\int d\xi u_{0}(\xi)$ is a geometric factor of
order unity, and $T_{\rm eff}(k,k')=\sum_{n}
\tilde{T}_{n,\ell}(k)\tilde{T}^{*}_{n,\rm
r}(k')/(\mu-\varepsilon_{\rm t}-\varepsilon_{n})$ is the effective
overlap integral [$\tilde{T}_{n,\ell/{\rm r}}(k)=\int dx
T_{\ell/{\rm r}}(x,k)|\psi_{n}(x)|^{2}$]. Below we will use
Eq.~(\ref{H4}) to analyze the magneto-conductance of the system
shown in Fig.~2. In doing so we assume that the overlap integral
$T_{\rm eff}(k,k')$ does not depend on the momenta $k$ and $k'$ .

It is clear from the effective Hamiltonian (\ref{H4}) that in the
presence of a magnetic field electronic transport through the SWNT
may be accompanied by inelastic processes. Although in principle
these may drive the vibration modes out of thermal equilibrium, we
will assume here that the coupling to the thermal bath is strong
enough for this not to happen. Under this condition the current
$I$ through the SWNT is to leading order in the tunneling coupling
given by the expression
\begin{eqnarray}\label{G0}
    I&=&G_{0}\sum_{n=0}^{\infty}\sum_{l=-n}^{\infty}P(n)|\langle
    n|e^{i\phi(\hat{b}^{\dag}+\hat{b})}|n+l\rangle|^{2} \times \\
    &&\int d\varepsilon\,\left[f_{\ell}(\varepsilon)(1-f_{\rm r}(\varepsilon-l\hbar\omega))
    -f_{\rm r}(\varepsilon)(1-f_{\ell}(\varepsilon-l\hbar\omega))\right]\,.\nonumber
\end{eqnarray}
Here $G_0=(2e/\hbar)\nu|T_{\rm eff}|^{2}$ is the zero-field
conductance,
$P(n)=(1-e^{-\beta\hbar\omega})e^{-n\beta\hbar\omega}$, where
$\beta\equiv1/k_{\rm B}T$, is the probability that the fundamental
mode is in state $|n\rangle$ with energy $n \hbar \omega$, and
$f_{\ell/\rm r}(\epsilon)\equiv(1+e^{\beta(\epsilon-\mu_{\ell/\rm
r})})^{-1}$ are Fermi distribution functions in the leads. The
restrictions on the tunneling channels imposed by the Pauli
principle are crucial; it easy to see from Eq.~(\ref{G0}) that if
all electron states in the right lead were empty it would follow
from the completeness of the set of vibronic states that there is
no effect of the magnetic field.

Finally, after integrating over the electron energy $\varepsilon$
in Eq.~(\ref{G0}) and using the completeness of the set of
vibronic states, {\em i.e.} $\sum_{n'}|\langle
n|e^{i\phi(\hat{b}^{\dag}+\hat{b}) }|n'\rangle|^{2}=1$, one
arrives at the final result for the linear conductance $G$,
\begin{eqnarray}\label{G1}
    \frac{G}{G_{0}}&=&
    \sum_{n=0}^{\infty}P(n)|\langle
    n|e^{i\phi(\hat{b}^{\dag}+\hat{b})}|n\rangle|^{2}+\\
    &&\sum_{n=0}^{\infty}\sum_{l=1}^{\infty}P(n)|\langle
    n|e^{i\phi(\hat{b}^{\dag}+\hat{b})}|n+l\rangle|^{2}
    \frac{2l\beta\hbar\omega}{e^{
    l\beta\hbar\omega}-1}\,.\nonumber
\end{eqnarray}
The first term in this expression gives the contribution to the
conductance from the elastic tunneling processes. The second term
is due to the inelastic processes, which contribute at finite
temperature.

Equation (\ref{G1}) has to be evaluated numerically, but the
asymptotic forms in the limits of high and low temperature,
\begin{equation}\label{G2}
    \frac{G}{G_{0}}=\left\{\begin{array}{l}
      1- \frac{1}{6}\frac{\hbar\omega}{k_{\rm B}T}\left(\frac{g4\pi Y_{0}LH}{\Phi_{0}}\right)^{2},\;\,
      \frac{\hbar\omega}{k_{\rm B}T}\ll 1,\, \phi \lesssim
      1 \\ \\
      e^{-\frac{1}{2}\left(\frac{g4\pi Y_{0}LH}{\Phi_{0}}\right)^{2}} ,\,\;\;\;\;\;\;\;\;\;\;\;\; \frac{\hbar\omega}{k_{\rm B}T}\gg 1
    \end{array}
    \right.
\end{equation}
follow readily from the identities $\langle
0|e^{i\phi(\hat{b}^{\dag}+\hat{b})}|0\rangle=e^{-\phi^{2}/2}$ and
$ \sum_{n´=0}^{\infty}(n'-n)|\langle
n|e^{i\phi(\hat{b}^{\dag}+\hat{b})}|n'\rangle|^{´2}=\phi^{2}$,
respectively.

It is instructive to express the probability amplitude
$A_{n}=\langle n|e^{i\phi(\hat{b}^{\dag}+\hat{b})}|n\rangle$ for
elastic transitions involving the $n$:th vibration mode in the
coordinate representation as
$$A_{n}=\int dy |\Phi_{n}(y)|^{2}e^{i\sqrt{2}\phi y}\,,$$
where $\Phi_{n}(y)$ is the wave function of the state $|n\rangle$.
From this expression it is evident that the main contribution to
the elastic tunneling amplitude corresponds to quasiclassical
trajectories passing through points where the probability to
``find" the SWNT is maximal, while electronic trajectories passing
through nodes of the SWNT wave functions do not contribute. This
can be interpreted as an effective multi-connectivity of the
system. We note, however, that this interpretation is strictly
valid only in the limit of vanishing temperature and bias voltage.

\begin{figure}
  \label{Fig3}
  \includegraphics[scale=0.7]{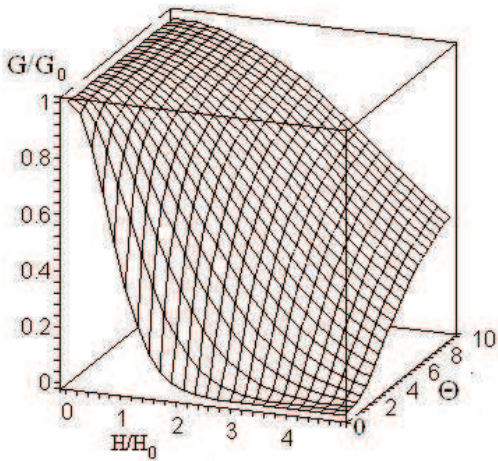}
  \vspace{-6 mm}
  \caption{ \label{fig:case22} Magneto-conductance of the doubly
  clamped SWNT shown in Fig.~2 as a function of normalized
  magnetic field and temperature calculated from Eq.~(\ref{G1})
  ($H_{0}=\Phi_{0}/g2\pi LY_{0}$) and $\Theta=k_{\rm B}T/\hbar\omega$).
  The
  negative magneto-conductance is due to a strong coupling
  between the coherent electrons and quantized flexural vibrations
  of the tube. This coupling leads to an effective multi-connectivity
  in the phase coherent electron transport and in a transverse magnetic
  field to destructive interference between different ``quantum"
  electron paths .
  }
  \vspace{-5 mm}
\end{figure}

Figure 3 
shows that quantum oscillations of a suspended SWNT result in a
positive magnetoresistance. The most striking feature of this
electronic conduction effect is its temperature dependence. It
comes from the dynamics of the entire nanotube, rather than from
the electron dynamics. Consequently, the characteristic
temperature scale of the resulting magnetoresistance --- which
decays as $T^{-1}$ for $T\gtrsim\hbar\omega$, where $\omega$ is
the ``swinging" frequency of the SWNT --- is anomalously small,
see Eq.~(\ref{G2}). A complete analysis has to account for the
strong nonlinearity of the magneto-conductance. As more and more
inelastic channels are switched on with increasing bias voltage
$V$, the influence of the magnetic field is suppressed in the same
way as by temperature. As a result the change in current due to a
magnetic field $H$ does not exceed $|G-G_0|\hbar\omega/e$ as $eV$
becomes larger than the vibration quantum $\hbar\omega$
\cite{restriction}. Hence, in order to easily detect the
magneto-conductance effect discussed here, one needs to be in the
ballistic transport regime \cite{TuMo}, where $G\simeq G_{Q}\equiv
2e^2/h \sim$~10$^{-4}$~mho. For a 1~$\mu$m long SWNT at $T=30$~mK
and $H \sim 20-40$~T we find from Eq.~(\ref{G2}) a relative
conductance change $|(G-G_0)/G_0|$ of about 1-3\%, which
corresponds to a maximal magneto-current $|I(H)-I(0)|$ of
0.1-0.3~pA. Its characteristic dependence on temperature and
transverse magnetic field makes this effect readily
distinguishable from the recently observed SWNT
magneto-conductance in a longitudinal field \cite{Dai}.

 In conclusion we have shown that quantum vibrations of a
current-carrying suspended single-wall carbon nanotube (SWNT)
subject to a transverse magnetic field couple electronic and
vibronic states -- forming what we call "swinging states" ---  in
such a way that a negative magneto-conductance can follow. This
effect may be used to detect nanomechanical SWNT vibrations in the
quantum regime. Other effects due to electron-phonon coupling can
be expected to occur in mesoscopic rings and in superconducting
Josephson junctions containing nanomechanical elements. A possible
consequence of the coupling to the macroscopic phase of a
superconductor is the prospect of using SQUIDs to detect
nanomechanical displacements in the quantum limit.

This work was supported in part by the Swedish VR, the Swedish
SSF, NSF grants DMR 02-37296 and DMR 04-39026, and by the European
Commission (EC) through project FP6-003673 CANEL of the IST
Priority. The views expressed in this publication are those of the
authors and not necessarily those of the EC.

\vspace{-5 mm}


\begin{thebibliography}
\bibliography{}

\bibitem{PhysicsToday}
K. C. Schwab and M. L. Roukes, Physics Today {\bf 58} (7) 36-42
(2005).

\bibitem{LaHaye}
M. D. LaHaye, O. Buu, B. Camarota and K. Schwab, Science {\bf
304}, 74 (2004); R. Knobel and A. N. Cleland, Nature {\bf 424},
291 (2003).

\bibitem{LeRoy}
B. J. LeRoy, S. G. Lemay, J. Kong and C. Dekker, Nature {\bf 432},
371, (2004).

\bibitem{OurNanoLett}
L. M. Jonsson, L. Y. Gorelik, R. I. Shekhter and M. Jonson, Nano
Lett. {\bf 5}, 1165 (2005).

\bibitem{et} In the expression for $\varepsilon_{\rm t}$ we omit
a term $\propto(\partial\hat{u}/\partial x)^{2}$ that gives rise
to an electron-vibron interaction at $H=0$, since it turns out not
to be important here.

\bibitem{Sav}
P. Werner, W. Zwerger, Europhys. Lett. {\bf 65}, 158 (2004);
S.~Savel'ev, F.~Nori, Phys. Rev. B {\bf 70}, 214415 (2004).

\bibitem{restriction}
Non-resonant tunneling requires $V$ and $T$ to be smaller than the
level spacing $\delta\varepsilon$ in a metallic SWNT (or smaller
than the energy gap $\Delta$ in a semiconducting nanotube). There
is, however, still room for a non-linear dependence of the current
on $V$ and $T$ in the interval $\hbar\omega \ll V \ll
\Delta,\delta\varepsilon$

\bibitem{TuMo}
Eq.~(10) was derived assuming $G \ll G_{\rm Q}\equiv 2e^2/h$, but
a more general analysis (to be published elsewhere) shows that
Eq.~¨(\ref{G2}) is still valid in the ballistic regime, $G\simeq
G_{\rm Q}$.

\bibitem{Dai}
J. Cao, Q. Wang, M. Rolandi, and H. Dai, Phys. Rev. Lett. {\bf
93}, 216803 (2004).

\end{thebibliography}
\end{document}